\documentclass[11pt,twocolumn,preprintnumbers,amsmath,amssymb,nofootinbib,superscriptaddress,aps,prd,numberedappendix]{openjournal}
\usepackage{natbib}
\usepackage{newtxtext,newtxmath}
\usepackage{amsmath,amssymb}
\usepackage{graphicx}
\usepackage{dcolumn}
\usepackage{bm}
\usepackage[dvipsnames]{xcolor}
\usepackage{xspace}
\usepackage{ulem}
\usepackage{lipsum}
\usepackage[colorlinks,linkcolor=red,citecolor=blue,urlcolor=blue ]{hyperref}
\usepackage{graphicx} 
\usepackage{multirow}
\usepackage{cancel}
\usepackage{yfonts}
\newcommand{\nv}{\hat{\bf n}}

\newcommand{\wtj}[6]{\left(\begin{array}{ccc} #1 & #2 & #3\\#4 & #5 & #6\end{array} \right)}
\newcommand{\mvec}[2]{\left(\begin{array}{c} #1 \\ #2 \end{array}\right)}
\newcommand{\mmat}[4]{\left(\begin{array}{cc} #1 & #2 \\ #3 & #4 \end{array}\right)}
\newcommand{\spt}[1]{\textswab{#1}}
\newcommand{\nmt}{{\tt NaMaster}\xspace}
\newcommand{\planck}{{\sl Planck}\xspace}
\newcommand{\mcmzz}[1]{M^{\bar{v}\bar{w} #1}_{\ell\ell'}}
\newcommand{\mcmzs}[2]{\textcolor{red}{M^{\bar{v}\spt{w}_{#1} #2}_{\ell\ell'}}}
\newcommand{\mcmsz}[2]{\textcolor{red}{M^{\spt{v}_{#1}\bar{w} #2}_{\ell\ell'}}}
\newcommand{\mcmss}[3]{\textcolor{blue}{M^{\spt{v}_{#1}\spt{w}_{#2} #3}_{\ell\ell'}}}

\begin{document}
\title{Pseudo-$C_\ell$s for spin-$s$ fields with component-wise weighting}

\author{David Alonso$^{1,*}$}
\email{$^*$david.alonso@physics.ox.ac.uk}
\affiliation{$^1$Department of Physics, University of Oxford, Denys Wilkinson Building, Keble Road, Oxford OX1 3RH, United Kingdom}

\date{\today}

\begin{abstract}
  We present a generalisation of the standard pseudo-$C_\ell$ approach for power spectrum estimation to the case of spin-$s$ fields weighted by a general positive-definite weight matrix that couples the different spin components of the field (e.g. $Q$ and $U$ maps in CMB polarisation analyses, or $\gamma_1$ and $\gamma_2$ shear components in weak lensing). Relevant use cases are, for example, data with significantly anisotropic noise properties, or situations in which different masks must be applied to the different field components. The weight matrix map is separated into a spin-0 part, which corresponds to the ``mask'' in the standard pseudo-$C_\ell$ approach, and a spin-$2s$ part sourced solely by the anisotropic elements of the matrix, leading to additional coupling between angular scales and $E/B$ modes. The general expressions for the mode-coupling coefficients involving the power spectra of these anisotropic weight components are derived and validated. The generalised algorithm is as computationally efficient as the standard approach. We implement the method in the public code \nmt.
\end{abstract}

\maketitle

\section{Introduction}\label{sec:intro}
  Angular power spectra are a centerpiece of a large fraction of analyses aimed at extracting information from the distribution of projected anisotropies. The study of the Cosmic Microwave Background (CMB) is perhaps the most popular field in which power spectra have played a central role \citep{astro-ph/9611174,astro-ph/9708203,astro-ph/0012120,1001.4635,1907.12875,2007.07289,2212.05642}, but their use has now become established in multiple other areas of observational cosmology, including the study of projected clustering and cosmic shear, as well as a myriad of other probes of the large-scale structure \citep{1809.09148,2010.09717,2304.00701}.
  
  The optimal and unbiased estimation of angular power spectra has therefore been a problem of primary importance for cosmology, and multiple methods have been developed to solve it. A manifestly optimal solution to the problem exists in the case of Gaussian random fields, the so-called ``quadratic maximum likelihood'' (or ``quadratic minimum-variance'') estimator (QML for short) \citep{astro-ph/9611174,astro-ph/9708203,astro-ph/0012120}. In this case, the field whose power spectrum is to be computed is first inverse-variance (IV) weighted (i.e. multiplied by an estimate of its inverse covariance matrix). The resulting IV-weighted map is then transformed to harmonic space, and its variance on different multipoles calculated. Finally, the impact of mode-coupling due to the IV weighting (described by the ``Fisher matrix'' in the nomenclature of \cite{astro-ph/9611174}) is estimated and corrected for. Unfortunately, the brute-force application of the QML estimator is often unfeasible (or, at least, impractical) \citep{astro-ph/0307515,1306.0005}. The method requires the computation and inversion of a large $N_{\rm pix}\times N_{\rm pix}$ covariance matrix, where $N_{\rm pix}$ is the number of pixels in the map. Although conjugate gradient methods and messenger field approaches may be used to avoid explicitly calculating, inverting, and storing this matrix \citep[e.g. ][]{1211.0585}, the computation of the IV-weighted map, and especially of the Fisher matrix, is usually very slow, particularly at high resolution.

  To sidestep this issue, approximate approaches have been devised that sacrifice optimality while preserving unbiasedness. This can be achieved through a judicious choice of covariance matrix with which to weight the field before estimating its spectrum. A particularly simple choice is to approximate the covariance matrix as being diagonal. Inverting such a matrix is trivial, and the IV-weighted map is a simple product between the original map and the diagonal of the inverse covariance. This is the basis of the pseudo-$C_\ell$ estimator \citep{astro-ph/0105302,astro-ph/0303414,astro-ph/0410394,1809.09603}. In this case, the diagonal of the inverse covariance is a ``mask'' that locally downweights or upweights particular sky areas (with completely masked areas interpreted as having infinite variance). Crucially, the impact of mode-coupling in the case of the pseudo-$C_\ell$ estimator (the analogue of the Fisher matrix for the QML) can be calculated analytically using fast methods. As per the assumption of a diagonal covariance, the pseudo-$C_\ell$ estimator is only close to optimal in the case of fields with a small correlation length, or for noise-dominated data in which the noise is close to white. These situations are commonly achieved, at least approximately, in multiple areas of cosmology, and thus the pseudo-$C_\ell$ estimator has become the tool of choice for most power spectrum-based analyses.

  A particular problem arises, however, in the case of spin-$s$ fields, i.e. vector and tensor fields on the sphere that transform non-trivially under rotations. Common examples are the CMB polarisation field and weak lensing shear (spin-$2$ fields), the lensing displacement and transverse peculiar velocities \citep{2305.15893} (spin-$1$ fields), or the polarisation of the stochastic gravitational wave background (spin-$4$) \citep{1608.06889}. In this case, although the noise properties of the field may still be well described as white (and thus amenable to the pseudo-$C_\ell$ approach), non-trivial anisotropy in the noise may lead to a differential variance in different field components, or to non-trivial correlations between them. Such a situation may occur, for example, in the case of CMB polarisation observations with insufficient cross-linking and unequal sensitivity across differently-oriented detectors, or it may be caused by artifacts in the images from which galaxy shapes are determined in cosmic shear analyses. In this case, according to the QML estimator, the optimal approach would be to weigh the field by its local inverse covariance prior to computing its harmonic-space variance. Unfortunately, the pseudo-$C_\ell$ estimator, as presented so far in the literature, assumes that all the components of a spin-$s$ field are weighted by the same weight map or mask. Depending on the level of anisotropy in the noise properties, this could lead to a significant degradation in the statistical uncertainties achieved by the estimator. Another potential motivation to apply anisotropic weights to a spin-$s$ field might be the desire to mask particular areas in the maps of one of the spin components due to the presence of systematics or strong foreground contamination that are not manifest in the other component.

  In this paper we present a generalisation of the standard pseudo-$C_\ell$ algorithm that is able to account for component-wise or ``anisotropic'' weights in spin-$s$ fields. The method is based on a decomposition of the weight matrix into its two irreducible representations: a spin-0 component that gives rise to the standard pseudo-$C_\ell$ coupling coefficients, and a spin-$2s$ component that incorporates the anisotropic nature of the weights, and which leads to additional couplings between different harmonic-space and spin modes. This is reminiscent of the results found in \cite{1401.2075} and \cite{1608.08833} when incorporating the impact of beam non-idealities in the pseudo-$C_\ell$ estimator for CMB polarisation data. Some of the results presented here were also introduced in \cite{2302.14656} when forward-modelling the effects of shear response corrections in weak lensing analyses.

  This paper is structured as follows. Section \ref{sec:meth} reviews the derivation of the standard pseudo-$C_\ell$ method and uses it to present the generalisation to the case of anisotropic weights (the main result of this paper). Section \ref{sec:val} validates the estimator by applying it to a variety of simulations incorporating different types of weight anisotropy. We then summarise our results and conclude in Section \ref{sec:conc}.

\section{Methods}\label{sec:meth}
  \subsection{The standard pseudo-$C_\ell$ algorithm}\label{ssec:meth.pcl}
    Let us start by briefly presenting the standard pseudo-$C_\ell$ estimator for spin-$s$ fields, highlighting the key steps that will enable us to derive the anisotropic estimator in the next section. The contents of this section are now common lore in the literature, and further details can be found in, e.g. \cite{astro-ph/0105302,astro-ph/0303414,astro-ph/0410394,1809.09603}. We repeat them here to facilitate the derivation of the generalised estimator presented in Section \ref{ssec:meth.ani}. Note that throughout this section, we use the convention used in {\tt HEALPix} \citep{astro-ph/0409513},  and {\tt libsharp} \citep{1303.4945} to define spin-$s$ fields and spin-weighted spherical harmonics.

    \subsubsection{Spin-$s$ fields and their power spectra}\label{sssec:meth.pcl.spins}
      Consider a spin-$s$ field with two components, $a_1(\nv)$ and $a_2(\nv)$. It is common to describe such fields in terms of a complex field $a\equiv a_1+i\,a_2$, for which the spin-$s$ transformation law can be written trivially: under a local rotation around the tangent plane by an angle $\psi$, the field transforms as $a\rightarrow a\,\exp(i\,s\psi)$. We can also consider the complex conjugate field $a^*=a_1-i\,a_2$ which, by construction is a spin-$(-s)$ field. The spectral decomposition of spin-$s$ fields is achieved by means of the spin-weighted spherical harmonic functions $_sY_{\ell m}(\nv)$ \citep{1967JMP.....8.2155G}. Specifically, the field $a$ and its spherical harmonic coefficients $_sa_{\ell m}$ are related via
     \begin{align}\label{eq:sht_s}
        &a(\nv)=-\sum_{\ell m}\,_sa_{\ell m}\,_sY_{\ell m}(\nv)\\
        &_sa_{\ell m}=-\int d\nv\,_sY^*_{\ell m}(\nv)\,a(\nv).
      \end{align}
      Likewise, the spherical harmonic coefficients of the complex field $a^*$ are constructed through a spin-$(-s)$ transform:
      \begin{align}
        &a^*(\nv)=-(-1)^s\sum_{\ell m}\,_{(-s)}(a^*)_{\ell m}\,_{(-s)}Y_{\ell m}(\nv)\\\label{eq:sht_s_b}
        &_{(-s)}(a^*)_{\ell m}\equiv -(-1)^s\int d\nv\,_{(-s)}Y^*_{\ell m}(\nv)\,a^*(\nv).
      \end{align}
      The factor $(-1)^s$ in the definition of $_{(-s)}(a^*)_{\ell m}$ is introduced to simplify the notation later on. Also, importantly, note that $_{(-s)}(a^*_{\ell m})$, defined in Eq. \ref{eq:sht_s_b}, is not the complex conjugate of $_sa_{\ell m}$. In fact, both coefficients are related via $_{(-s)}(a^*_{\ell m})=(-1)^m(_sa_{\ell -m})^*$. Finally, it is common to express the harmonic coefficients of a spin-$s$ field in terms of the coefficients of a scalar $E$-mode field $a^E_{\ell m}$, and a pseudo-scalar $B$-mode $a^B_{\ell m}$, defined as
      \begin{equation}
        _sa_{\ell m}\equiv a^E_{\ell m}+ia^B_{\ell m},\hspace{6pt}
        _{(-s)}(a^*)_{\ell m}\equiv a^E_{\ell m}-ia^B_{\ell m}.
      \end{equation}

      Consider now two fields $a$ and $b$ with spins $s_a$ and $s_b$. The power spectrum between these two fields is defined as
      \begin{equation}\label{eq:clab}
        \langle\,_{s_a}a_{\ell m}\,(_{s_b}b_{\ell m})^*\rangle\equiv\delta_{\ell\ell'}\delta_{mm'}\,C^{ab}_\ell.
      \end{equation}
      Likewise, we may calculate the power spectrum between $a$ and the complex conjugate of $b$, for which we will use the shorthand $C^{a\bar{b}}_\ell$:
      \begin{equation}
        \langle\,_{s_a}a_{\ell m}\,(_{(-s_b)}(b^*)_{\ell m})^*\rangle\equiv\delta_{\ell\ell'}\delta_{mm'}\,C^{a\bar{b}}_\ell.
      \end{equation}
      The complex-valued power spectra $C^{ab}_\ell$ and $C^{a\bar{b}}_\ell$ may be expressed in terms of the real-valued power spectra of the scalar $E$- and $B$-modes:
      \begin{align}\nonumber
        &C^{ab}_\ell=C^{EE}_\ell+C^{BB}_\ell+i(C^{BE}_\ell-C^{EB}_\ell)\\\label{eq:clab2eb}
        &C^{a\bar{b}}_\ell=C^{EE}_\ell-C^{BB}_\ell+i(C^{BE}_\ell+C^{EB}_\ell).
      \end{align}
      Or, in other words, we may reconstruct the $E$ and $B$ power spectra from $C^{ab}_\ell$ and $C^{a\bar{b}}_\ell$:
      \begin{align}
        &C^{EE}_\ell=\frac{1}{2}{\rm Re}(C^{ab}_\ell+C^{a\bar{b}}_\ell),\hspace{6pt}
        C^{BB}_\ell=\frac{1}{2}{\rm Re}(C^{ab}_\ell-C^{a\bar{b}}_\ell),\\
        &C^{BE}_\ell=\frac{1}{2}{\rm Im}(C^{ab}_\ell+C^{a\bar{b}}_\ell),\hspace{6pt}
        C^{EB}_\ell=\frac{1}{2}{\rm Im}(C^{a\bar{b}}_\ell-C^{ab}_\ell).
      \end{align}

    \subsubsection{Pseudo-$C_\ell$s}\label{sssec:meth.pcl.pcl}
      In more realistic situations, we often deal with weighted or ``masked'' fields: $\tilde{a}(\nv)\equiv v(\nv)\,a(\nv)$, where $v(\nv)$ is a real-valued scalar map normally called the ``mask''. As discussed in the introduction, $v$ serves not just as a pure mask, removing particular sky areas where $v=0$, but should in general be thought of as a local weighting function, the design of which affects the statistical uncertainties of any measurements derived from the masked field. Ideally, $v(\nv)$ would correspond to an inverse-variance weight, with masked areas corresponding to regions with a very large (infinite) variance. By weighting our spin-$s$ field with a scalar mask, we are thus implicitly assuming that the variance of both spin components ($a_1$ and $a_2$) is the same or, in general, that their local covariance matrix is proportional to the identity. Generalising this assumption will be the subject of Section \ref{ssec:meth.ani}.

      The multiplication by $v$ leads to coupling between the different angular scales of the field. Specifically, the harmonic coefficients of the masked field and the unmasked one are related via
      \begin{equation}\label{eq:conv_s}
        _s\tilde{a}_{\ell m}=\sum_{\substack{\ell' m'\\\ell''m''}}v_{\ell'' m''}\,_sa_{\ell' m'}\,{\cal G}^{-s\,s\,0}_{\ell m,\ell'm',\ell''m''},
      \end{equation}
      where $v_{\ell m}$ are the spherical harmonic coefficiens of the mask\footnote{Note that these are defined as $v(\nv)=\sum_{\ell m} Y_{\ell m}(\nv)\,v_{\ell m}$, where $Y_{\ell m}$ are the standard spherical harmonic function. In particular, note the absence of the $-1$ sign that appears in the spin-$s$ spherical harmonic transforms (Eq. \ref{eq:sht_s}). This is a mostly inconvenient convention that often leads to significant confusion.}, and we have defined the \emph{Gaunt coefficients}:
      \begin{equation}\label{eq:gaunt}
        {\cal G}^{s_1s_2s_3}_{\ell m,\ell'm',\ell''m''}\equiv\int d\nv\,_{-s_1}Y^*_{\ell m}(\nv)\,_{s_2}Y_{\ell'm'}(\nv)\,_{s_3}Y_{\ell''m''}(\nv).
      \end{equation}
      The Gaunt coefficients may be expressed in terms of the Wigner-$3j$ symbols as:
      \begin{align}\nonumber
        {\cal G}^{s_1s_2s_3}_{\ell m,\ell'm',\ell''m''}=&(-1)^{s_1+m}\sqrt{\frac{(2\ell+1)(2\ell'+1)(2\ell''+1)}{4\pi}}\\
        &\times\wtj{\ell}{\ell'}{\ell''}{-m}{m'}{m''}
        \wtj{\ell}{\ell'}{\ell'}{-s_1}{-s_2}{-s_3}.
      \end{align}
      A similar relation may be found for the harmonic coefficients of the complex field:
      \begin{equation}
        _{(-s)}(\tilde{a}^*)_{\ell m}=\sum_{\substack{\ell' m'\\\ell''m''}}(-1)^{\sum_\ell}v_{\ell'' m''}\,_{(-s)}(a^*)_{\ell' m'}\,{\cal G}^{-s\,s\,0}_{\ell m,\ell'm',\ell''m''},
      \end{equation}
      where $\sum_\ell\equiv\ell+\ell'+\ell''$, and we have made use of the following symmetry of the $3j$ symbols:
      \begin{equation}
        \wtj{\ell}{\ell'}{\ell''}{-s_1}{-s_2}{-s_3}=(-1)^{\sum_\ell}\wtj{\ell}{\ell'}{\ell''}{s_1}{s_2}{s_3}.
      \end{equation}

      Let us now define the pseudo-$C_\ell$ of two masked fields $(\tilde{a},\tilde{b})$ with masks $(v,w)$, and spins $(s_a,s_b)$:
      \begin{equation}\label{eq:pcldef_gen}
        \tilde{C}^{ab}_\ell\equiv\frac{1}{2\ell+1}\sum_{m}\,_{s_a}\tilde{a}_{\ell m}\,_{s_b}\tilde{b}^*_{\ell m}
      \end{equation}
      (and likewise for the power spectrum involving the complex of $b$, $\tilde{C}^{a\bar{b}}_\ell$). Taking the expectation value of the equation above, after expanding the masked fields using Eq. \ref{eq:conv_s}, and using Eq. \ref{eq:clab} to introduce the power spectrum of the unmasked fields, we obtain:
      \begin{align}\nonumber
        \langle\tilde{C}^{ab}_\ell\rangle=
        &\sum_{\substack{\ell'\ell''m''\\\ell_3m_3}}v_{\ell''m''}w^*_{\ell_3m_3}C^{ab}_{\ell'}\\
        &\hspace{6pt}\times\frac{1}{2\ell+1}\sum_{mm'}{\cal G}^{-s_as_a0}_{\ell m,\ell'm',\ell''m''}{\cal G}^{-s_bs_b0}_{\ell m,\ell'm',\ell_3m_3}.
      \end{align}
      The sum over Gaunt coefficients can be simplified using the orthogonality relation for the $3j$ symbols
      \begin{equation}
        \sum_{mm'}\wtj{\ell}{\ell'}{\ell''}{m}{m'}{m''}\wtj{\ell}{\ell'}{\ell_3}{m}{m'}{m_3}=\frac{\delta_{\ell''\ell_3}\delta_{m''m_3}}{2\ell''+1}.
      \end{equation}
      This allows for further simplifications, yielding the final result
      \begin{equation}\label{eq:pcl_ab}
        \langle\tilde{C}^{ab}_\ell\rangle=\sum_{\ell'}M^{vw}_{\ell\ell'}\,C^{ab}_{\ell'},
      \end{equation}
      where the mode-coupling matrix $M^{vw}_{\ell\ell'}$ is
      \begin{align}\nonumber
        M^{vw}_{\ell\ell'}\equiv
        &(-1)^{s_a+s_b}\sum_{\ell''}\frac{(2\ell'+1)(2\ell''+1)}{4\pi}\\\label{eq:mcm_std}
        &\hspace{6pt}\times\tilde{C}^{vw}_{\ell''}\wtj{\ell}{\ell'}{\ell''}{s_a}{-s_a}{0}\wtj{\ell}{\ell'}{\ell''}{s_b}{-s_b}{0},
      \end{align}
      and $\tilde{C}^{vw}_\ell$ is the pseudo-$C_\ell$ of the two masks.
      
      Following the same steps one can derive the a similar relation for the power spectrum involving the complex conjugate field:
      \begin{equation}\label{eq:pcl_abbar}
        \langle\tilde{C}^{a\bar{b}}_\ell\rangle=\sum_{\ell'}\bar{M}^{vw}_{\ell\ell'}\,C^{a\bar{b}}_{\ell'},
      \end{equation}
      with the modified coupling matrix
      \begin{align}\nonumber
        \bar{M}^{vw}_{\ell\ell'}\equiv
        &(-1)^{s_a+s_b}\sum_{\ell''}\frac{(2\ell'+1)(2\ell''+1)}{4\pi}(-1)^{\sum_\ell}\\\label{eq:mcm_std_b}
        &\hspace{6pt}\times\tilde{C}^{vw}_{\ell''}\wtj{\ell}{\ell'}{\ell''}{s_a}{-s_a}{0}\wtj{\ell}{\ell'}{\ell''}{s_b}{-s_b}{0}.
      \end{align}
      From equations \ref{eq:pcl_ab} and \ref{eq:pcl_abbar}, and using the relation between these power spectra and the $E$- and $B$-mode spectra (Eq. \ref{eq:clab2eb}), we can derive expressions for the pseudo-$C_\ell$ of the latter. The final result is:
      \begin{equation}
        \langle\tilde{C}^{\alpha\beta}_\ell\rangle=\sum_{\ell'\alpha'\beta'}M^{\alpha\beta,\alpha'\beta'}_{\ell\ell'}C^{\alpha'\beta'}_{\ell'},
      \end{equation}
      where the Greek symbols ($\alpha$, $\beta$, etc.) stand for the $E$ and $B$ components, and the mode-coupling coefficients are
      \begin{align}
        &M^{EE,EE}_{\ell\ell'}=M^{EB,EB}_{\ell\ell'}=M^{BE,BE}_{\ell\ell'}=M^{BB,BB}_{\ell\ell'}=M^{vw+}_{\ell\ell'},\\\nonumber
        &M^{EE,BB}_{\ell\ell'}=-M^{EB,BE}_{\ell\ell'}=-M^{BE,EE}_{\ell\ell'}=M^{BB,EE}_{\ell\ell'}=M^{vw-}_{\ell\ell'},
      \end{align}
      with all other coefficients, which correspond to parity-odd couplings such as $M^{EE,EB}_{\ell\ell'}$, equal to zero. Above we have defined
      \begin{align}\nonumber
        M^{vw,\pm}_{\ell\ell'}&\equiv\frac{1}{2}(M^{vw}_{\ell\ell'}\pm\bar{M}^{vw}_{\ell\ell'})\\\nonumber
        &=(-1)^{s_a+s_b}\sum_{\ell''}\frac{(2\ell'+1)(2\ell''+1)}{4\pi}\frac{1\pm(-1)^{\sum_\ell}}{2}\\
        &\hspace{12pt}\times\tilde{C}^{vw}_{\ell''}\wtj{\ell}{\ell'}{\ell''}{s_a}{-s_a}{0}\wtj{\ell}{\ell'}{\ell''}{s_b}{-s_b}{0}.
      \end{align}

      Having derived the relation between $\tilde{C}^{\alpha\beta}_\ell$ and $C^{\alpha\beta}_\ell$, an unbiased estimator of the power spectrum can be constructed by inverting the mode-coupling matrix. In the cases where the mode-coupling matrix is not invertible (e.g. for small sky patches), it is common to bin the power spectrum into ``bandpowers'' ($\ell$-bins) and invert the better-behaved binned coupling matrix. In these cases it is important to make sure that any residual mode-coupling left by this approximate inversion is propagated when interpreting the measured power spectra. The details of these subsequent operations are described in e.g. \cite{1809.09603}, and will not be repeated here.
    
  \subsection{Anisotropic weighting of spin-$s$ fields}\label{ssec:meth.ani}
    Consider now the more general case of observations of a spin-$s$ field with anisotropic noise properties. To begin with, and for clarity, we will make use of a vectorial notation, writing the field $a$ as a vector with components ${\bf a}\equiv(a_1,a_2)$. In this case, the noise variance of the two spin components is not necessarily the same, and there may be a non-zero correlation between them. If we assume the noise to still be spatially white (i.e. uncorrelated between pixels), the optimally-weighted field in this case would be
    \begin{equation}\label{eq:weight_gen_vec}
      \tilde{\bf a}\equiv\mvec{\tilde{a}_1}{\tilde{a}_2}_{\nv}\equiv\mmat{v_{11}}{v_{12}}{v_{12}}{v_{22}}_{\nv} \mvec{a_1}{a_2}_{\nv}\equiv {\sf V}\cdot{\bf a},
    \end{equation}
    where $v_{ij}$ are the elements of the inverse-noise covariance, and we have explicitly noted that both $a_i$ and $v_{ij}$ are spatially-dependent (although only $a_i$ is a stochastic field).

    To make further progress, let us consider the transformation laws of ${\bf a}$ and ${\sf V}$ under a rotation by angle $\psi$ around an axis perpendicular to the tangent plane. Since ${\bf a}$ is a spin-$s$ field:
    \begin{align}
      &{\bf a}\rightarrow {\sf R}_{s\psi}\cdot{\bf a}\\\label{eq:transf_V}
      &{\sf V}\rightarrow {\sf R}_{s\psi}\cdot{\sf V}\cdot{\sf R}_{s\psi}^T,
    \end{align}
    where $R_\varphi$ is a rotation matrix by angle $\varphi$:
    \begin{equation}
      {\sf R}_\varphi=\mmat{\cos\varphi}{-\sin\varphi}{\sin\varphi}{\cos\varphi}.
    \end{equation}
    Thus, as could be expected, the weighted field is also spin-$s$: $\tilde{\bf a}\rightarrow{\sf R}_{s\psi}\cdot\tilde{\bf a}$. We can now separate ${\sf V}$ into its irreducible representations under 2D rotations. Using the complex number notation introduced in Section \ref{sssec:meth.pcl.spins}, let us define two fields
    \begin{equation}
      \bar{v}\equiv\frac{v_{11}+v_{22}}{2},\hspace{12pt}\spt{v}\equiv\frac{v_{11}-v_{22}}{2}+i\,v_{12}.
    \end{equation}
    It is straightforward to show that $\bar{v}$ (the average weight of the two spin components) is a spin-0 field, and that $\spt{v}$, which collects all the anisotropic properties of the weight matrix, is a \emph{spin-$2s$} field. Furthermore, we can now express the generalised weighting of Eq. \ref{eq:weight_gen_vec} fully in complex form as:
    \begin{equation}
      \tilde{a}(\nv)=\bar{v}(\nv)\,a(\nv)+\spt{v}(\nv)\,a^*(\nv).
    \end{equation}
    The first term is equivalent to the standard isotropic weighting described in Section \ref{sssec:meth.pcl.pcl}. In turn, the second term incorporates the full impact of the anisotropic weighting and is exactly zero for the standard isotropic masking approach (since $\spt{v}=0$ in that case). As a final sanity check, since $a^*$ is a spin-$(-s)$ field, and $\spt{v}$ is a spin-$2s$ field, the product $\spt{v}\,a^*$ is spin-$s$, as expected.

    It is important to point out that this result is general, regardless of the nature of the weight matrix ${\sf V}$. The masked field $\tilde{\bf a}\equiv{\sf V}\cdot{\bf a}$ must retain its original spin nature after masking, and therefore:
    \begin{equation}
      \tilde{\bf a}'={\sf R}_{s\psi}\cdot\tilde{\bf a}={\sf R}_{s\psi}\cdot{\sf V}\cdot{\bf a},
    \end{equation}
    where primed $(')$ quantities are transformed under a rotation by angle $\psi$. At the same time:
    \begin{equation}
      \tilde{\bf a}'={\sf V}'\cdot{\bf a}'={\sf V}'\cdot{\sf R}_{s\psi}\cdot{\bf a}.
    \end{equation}
    This can only hold if ${\sf V}$ transforms as in Eq. \ref{eq:transf_V}, even if it was not constructed from a inverse covariance matrix.
    
  \subsection{Generalised pseudo-$C_\ell$s}\label{ssec:meth.genpcl}
    We now incorporate the impact of anisotropic weighting in the pseudo-$C_\ell$ of two fields $(\tilde{a},\tilde{b})$, with weights $(\bar{v},\spt{v})$ and $(\bar{w},\spt{w})$, and spins $(s_a,s_b)$. We start by writing out the harmonic coefficients of one of these weighted fields:
    \begin{align}
      \,_s\tilde{a}_{\ell m}=\sum_{\ell' m'}\sum_{\ell'' m''}&\bar{v}_{\ell'' m''}\,_sa_{\ell'm'}\,{\cal G}^{-s_a,s_a,0}_{\ell m,\ell'm',\ell''m''}\\\nonumber
      &-(-1)^{s_a}\,_{2s}\spt{v}_{\ell''m''}\,_{-s}(a^*)_{\ell' m'}\,{\cal G}^{-s_a,-s_a,2s_a}_{\ell m,\ell'm',\ell''m''},
    \end{align}
    where the Gaunt coefficients were defined in Eq. \ref{eq:gaunt}. Comparing this structure with Eqs. \ref{eq:conv_s} and \ref{eq:mcm_std}, we can see that the anisotropic weights will give rise to additional terms in the mode-coupling coefficients involving Wigner-$3j$ symbols with spins combinations $(s,s,-2s)$.
    
    The derivation of these terms, although lengthier, follows the same steps as those described in Section \ref{ssec:meth.pcl} for the standard pseudo-$C_\ell$ coupling matrix, and will not be repeated here. The result is:
    \begin{widetext}
    \begin{align}\nonumber
      &M^{EE,EE}_{\ell\ell'}=\mcmzz{+}-\mcmzs{E}{+}-\mcmsz{E}{+}+\mcmss{E}{E}{+}+\mcmss{B}{B}{-},
      \hspace{12pt}
      M^{EB,EE}_{\ell\ell'}=-\mcmzs{B}{+}+\mcmsz{B}{-}+\mcmss{E}{B}{+}-\mcmss{B}{E}{-},
      \\\nonumber
      &M^{EE,EB}_{\ell\ell'}=-\mcmzs{B}{+}-\mcmsz{B}{-}+\mcmss{E}{B}{+}-\mcmss{B}{E}{-},
      \hspace{12pt}
      M^{EB,EB}_{\ell\ell'}=\mcmzz{+}+\mcmzs{E}{+}-\mcmsz{E}{+}-\mcmss{E}{E}{+}-\mcmss{B}{B}{-},
      \\\nonumber
      &M^{EE,BE}_{\ell\ell'}=-\mcmzs{B}{-}-\mcmsz{B}{+}+\mcmss{B}{E}{+}-\mcmss{E}{B}{-},
      \hspace{12pt}
      M^{EB,BE}_{\ell\ell'}=-\mcmzz{-}+\mcmzs{E}{-}-\mcmsz{E}{-}+\mcmss{E}{E}{-}+\mcmss{B}{B}{+},
      \\\nonumber
      &M^{EE,BB}_{\ell\ell'}=\mcmzz{-}+\mcmzs{E}{-}+\mcmsz{E}{-}+\mcmss{E}{E}{-}+\mcmss{B}{B}{+},
      \hspace{12pt}
      M^{EB,BB}_{\ell\ell'}=\mcmzs{B}{-}-\mcmsz{B}{+}+\mcmss{E}{B}{-}-\mcmss{B}{E}{+},
      \\\nonumber
      &M^{BB,EE}_{\ell\ell'}=\mcmzz{-}-\mcmzs{E}{-}-\mcmsz{E}{-}+\mcmss{E}{E}{-}+\mcmss{B}{B}{+}
      \hspace{12pt}
      M^{BE,EE}_{\ell\ell'}=\mcmzs{B}{-}-\mcmsz{B}{+}+\mcmss{B}{E}{+}-\mcmss{E}{B}{-},
      \\\nonumber
      &M^{BB,EB}_{\ell\ell'}=-\mcmzs{B}{-}-\mcmsz{B}{+}+\mcmss{E}{B}{-}-\mcmss{B}{E}{+},
      \hspace{12pt}
      M^{BE,EB}_{\ell\ell'}=-\mcmzz{-}-\mcmzs{E}{-}+\mcmsz{E}{-}+\mcmss{E}{E}{-}+\mcmss{B}{B}{+},
      \\\nonumber
      &M^{BB,BE}_{\ell\ell'}=-\mcmzs{B}{+}-\mcmsz{B}{-}+\mcmss{B}{E}{-}-\mcmss{E}{B}{+},
      \hspace{12pt}
      M^{BE,BE}_{\ell\ell'}=\mcmzz{+}-\mcmzs{E}{+}+\mcmsz{E}{+}-\mcmss{E}{E}{+}-\mcmss{B}{B}{-},
      \\\label{eq:mcm_gen}
      &M^{BB,BB}_{\ell\ell'}=\mcmzz{+}+\mcmzs{E}{+}+\mcmsz{E}{+}+\mcmss{E}{E}{+}+\mcmss{B}{B}{-},
      \hspace{12pt}
      M^{BE,BB}_{\ell\ell'}=-\mcmzs{B}{+}+\mcmsz{B}{-}-\mcmss{E}{B}{+}+\mcmss{B}{E}{-},
    \end{align}
    \end{widetext}
    where we have defined mode-coupling matrices involving the spin-0 masks, and the $E/B$ components of the spin-$2s$ masks. For clarity, we have coloured the matrices involving one spin-0 mask and one spin-$2s$ mask in red, and the matrices involving two spin-$2s$ masks in blue. These matrices are
    \begin{widetext}
    \begin{align}
      &M^{\bar{v}\bar{w}\pm}_{\ell\ell'}\equiv(-1)^{s_a+s_b}\sum_{\ell''}\frac{(2\ell'+1)(2\ell''+1)}{4\pi}\frac{1\pm(-1)^{\sum_\ell}}{2}\tilde{C}^{\bar{v}\bar{w}}_{\ell''}\wtj{\ell}{\ell'}{\ell''}{s_a}{-s_a}{0}\wtj{\ell}{\ell'}{\ell''}{s_b}{-s_b}{0},\\
      &M^{\bar{v}\spt{w}_\alpha\pm}_{\ell\ell'}\equiv(-1)^{s_a}\sum_{\ell''}\frac{(2\ell'+1)(2\ell''+1)}{4\pi}\frac{1\pm(-1)^{\sum_\ell}}{2}\tilde{C}^{\bar{v}\spt{w}_{\alpha}}_{\ell''}\wtj{\ell}{\ell'}{\ell''}{s_a}{-s_a}{0}\wtj{\ell}{\ell'}{\ell''}{s_b}{s_b}{-2s_b},\\
      &M^{\spt{v}_\alpha\bar{w}\pm}_{\ell\ell'}\equiv(-1)^{s_b}\sum_{\ell''}\frac{(2\ell'+1)(2\ell''+1)}{4\pi}\frac{1\pm(-1)^{\sum_\ell}}{2}\tilde{C}^{\spt{v}_\alpha\bar{w}}_{\ell''}\wtj{\ell}{\ell'}{\ell''}{s_a}{s_a}{-2s_a}\wtj{\ell}{\ell'}{\ell''}{s_b}{-s_b}{0},\\
      &M^{\spt{v}_\alpha\spt{w}_\beta\pm}_{\ell\ell'}\equiv\sum_{\ell''}\frac{(2\ell'+1)(2\ell''+1)}{4\pi}\frac{1\pm(-1)^{\sum_\ell}}{2}\tilde{C}^{\spt{v}_\alpha\spt{w}_\beta}_{\ell''}\wtj{\ell}{\ell'}{\ell''}{s_a}{s_a}{-2s_a}\wtj{\ell}{\ell'}{\ell''}{s_b}{s_b}{-2s_b}.
    \end{align}
    \end{widetext}
    A simpler expression that summarises this result is presented in Appendix \ref{app:simpler}
    
    We can see that anisotropic weighting leads to a much more complex structure of the mode-coupling matrix. The non-zero coupling coefficients present in the standard pseudo-$C_\ell$ receive additional contributions from the parity-even power spectra of the spin-$2s$ mask (e.g. $\tilde{C}^{\bar{v}\spt{w}_E}_\ell$ or $\tilde{C}^{\spt{v}_B\spt{w}_B}_\ell$), and the parity-odd mask spectra (e.g. $\tilde{C}^{\spt{v}_E\spt{w}_B}_\ell$ or $\tilde{C}^{\bar{v}\spt{w}_B}_\ell$) excite the parity-odd elements of the mode-coupling matrix that were previously zero (e.g. $M^{EE,EB}_{\ell\ell'}$). As another basic sanity check, we can verify that, when both masks are identical, many of the terms above become equivalent (e.g. $M_{\ell\ell'}^{EE,EB}$ and $M_{\ell\ell'}^{EE,BE}$).

    In spite of the additional structure, anisotropic weighting does not increase significantly the computational cost of calculating the mode-coupling matrix. The bottleneck here is usually the calculation of the Wigner-$3j$ symbols and, in this case, only one additional spin combination (with $(s,s,-2s)$ in the bottom row) must be computed. This is similar to the case of purification of spin-2 fields \citep{astro-ph/0511629,0903.2350,1809.09603}, where two additional sets of $3j$ coefficients must be calculated. The task of estimating power spectra in the presence of anisotropic weights is thus as simple as in the standard scenario.

    Note that, although we have provided the expressions above for the most general case of two non-scalar fields, both with anisotropic weights, the form of these coupling coefficients for simpler combinations is straightforward to derive. If one of the fields has isotropic weights, all terms involving the spin-$2s$ mask of that field vanish. If one of the fields is a scalar, one must further discard all coupling terms involving the $B$-modes of that field (e.g. $M^{EE,BB}_{\ell\ell'}$), and simply interpret its $E$-modes as its standard harmonic coefficients.

    The formalism we have just described is implemented in the public pseudo-$C_\ell$ code \nmt\footnote{\url{https://github.com/LSSTDESC/NaMaster}.} \citep{1809.09603}. Validating this implementation is the subject of the next section.

\section{Validation}\label{sec:val}
  \subsection{Simple anisotropic weight models}\label{ssec:val.simp}
    \begin{figure*}
        \centering
        \includegraphics[width=0.40\textwidth]{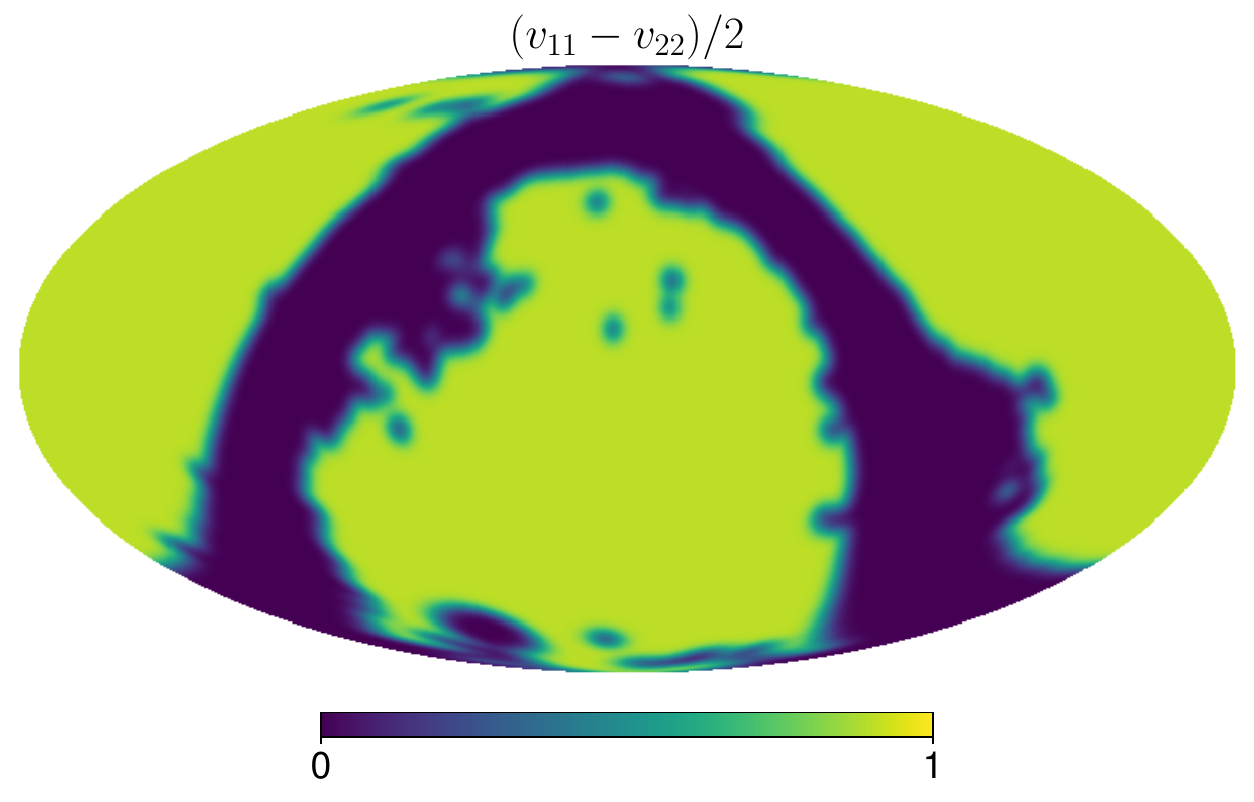}
        \includegraphics[width=0.40\textwidth]{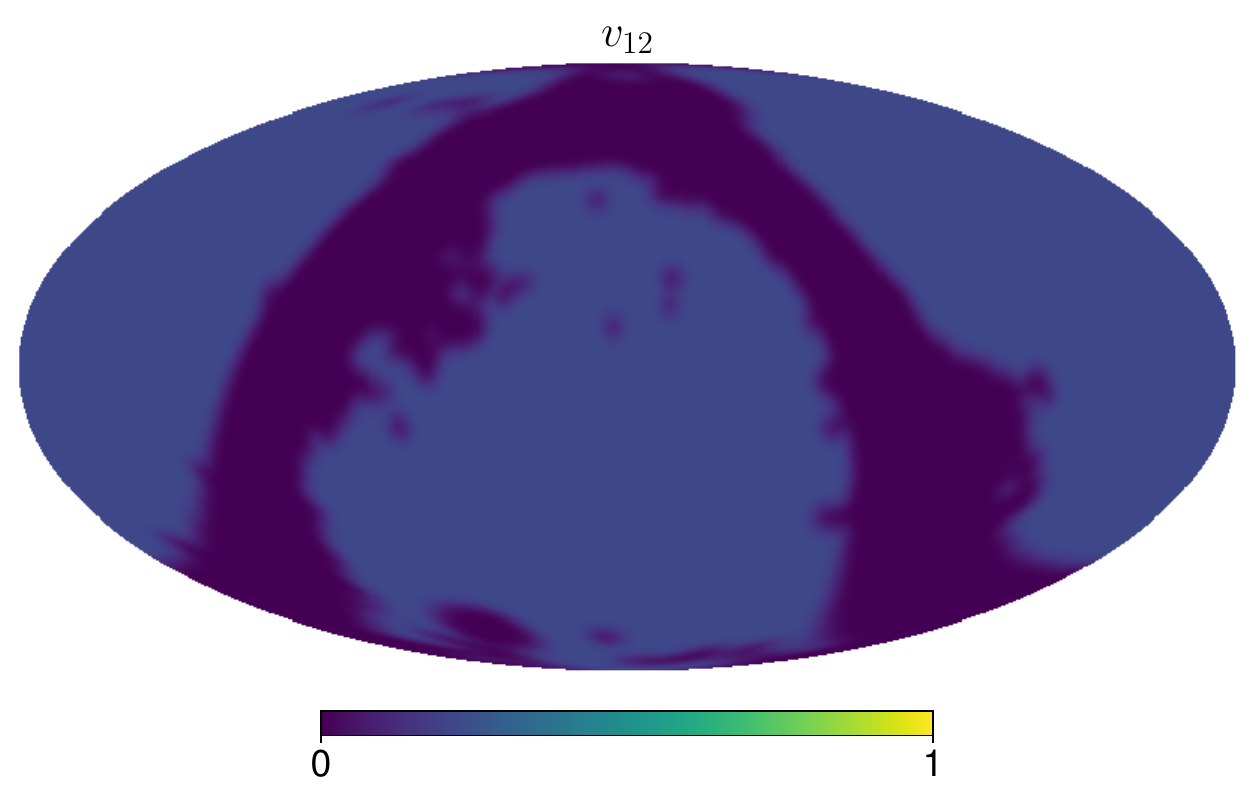}
        \includegraphics[width=0.33\textwidth]{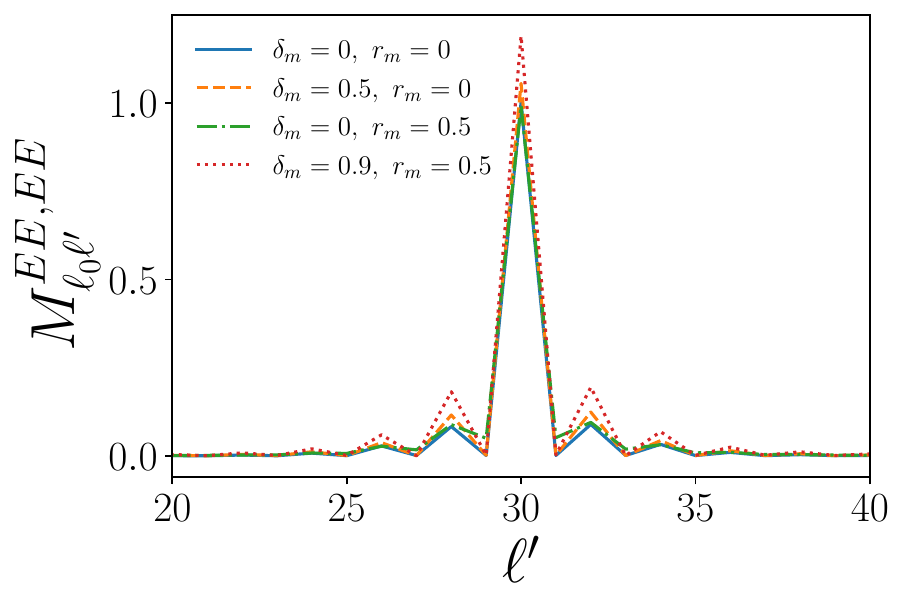}
        \includegraphics[width=0.33\textwidth]{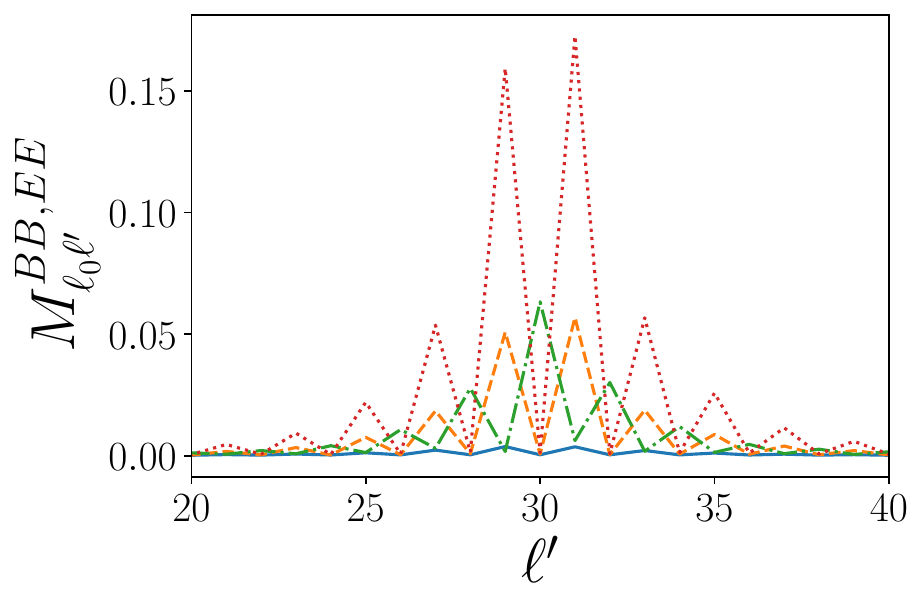}
        \includegraphics[width=0.33\textwidth]{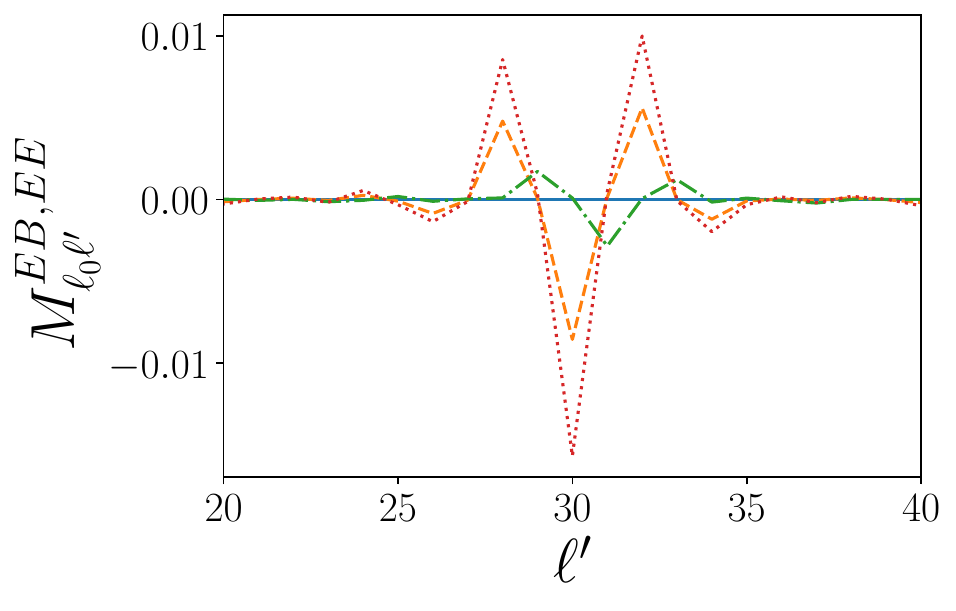}
        \caption{{\sl Top row:} elements of the anisotropic weighting matrix for the first validation set with $\delta_m=0.9$ and $r_m=0.5$. {\sl Bottom row:} Rows of the mode-coupling matrix describing the leakage of the $EE$ power spectrum onto the $EE$, $BB$, and $EB$ power spectra (left, center, and right panels, respectively) at multipole $\ell_0=30$, for different levels of weight anisotropy described by the parameters $(\delta_m,r_m)$.}
        \label{fig:wmat}
    \end{figure*}
    \begin{figure}
      \centering
      \includegraphics[width=0.49\textwidth]{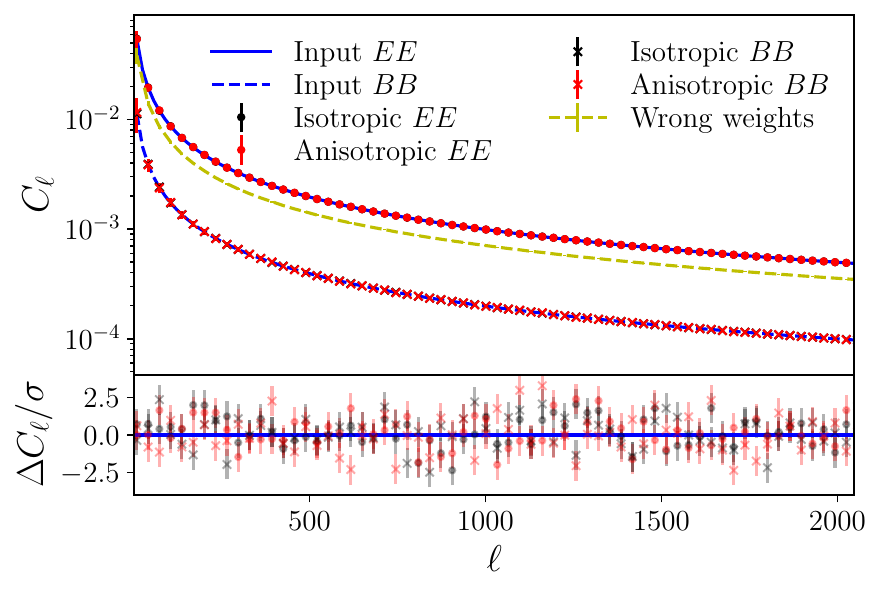}
      \caption{Angular power spectra from our first validation suite. The input $EE$ and $BB$ power spectra are shown in solid and dashed blue lines, respectively. The mean spectra obtained from 100 simulations with isotropic weights are shown as black markers with error bars corresponding to the scatter in the mean (i.e. the scatter over simulations divided by $\sqrt{N_{\rm sim}}$). The red markers show the results for simulations with anisotropic weights. The bottom panel shows the residuals with respect to the input power spectra. The anisotropic pseudo-$C_\ell$ estimator is unbiased (as is the standard isotropic estimator, unsurprisingly). The yellow line shows the result of computing the $BB$ power spectrum of anisotropically-weighted simulations assuming isotropic weights.}
      \label{fig:clval1}
    \end{figure}
    \begin{figure}
      \centering
      \includegraphics[width=0.49\textwidth]{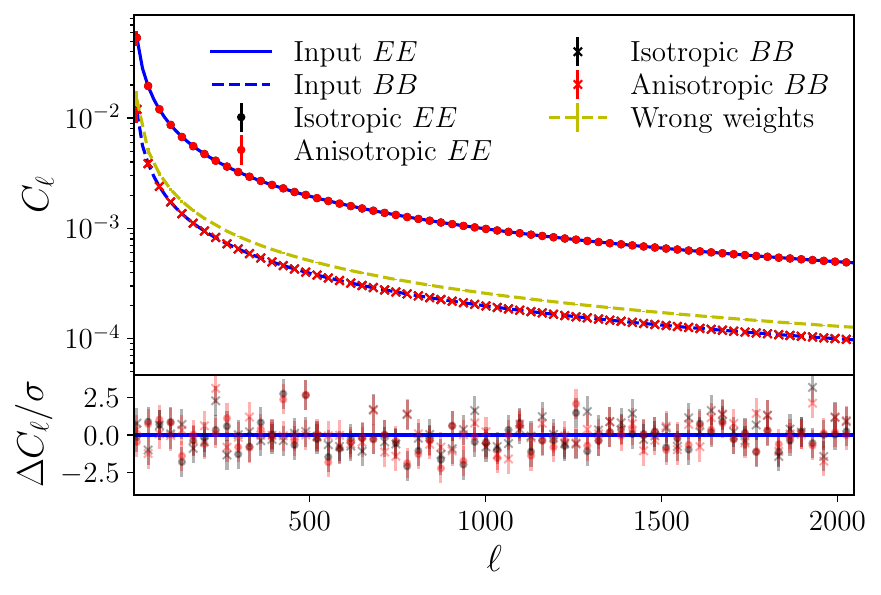}
      \caption{As Figure \ref{fig:clval1} for our second validation set, in which masks covering different sky areas are applied to the two spin components of the field. See caption of Fig. \ref{fig:clval1} for details.}
      \label{fig:clval2}
    \end{figure}
    To validate the estimator derived in Section \ref{ssec:meth.genpcl} in the presence of anisotropic weights, we make use of two sets of idealised simulations. Both sets are based on full-sky realisations of a spin-$2$ field with power spectra
    \begin{equation}
      C_\ell^{EE}=\frac{1}{\ell+\ell_*},\hspace{6pt}
      C_\ell^{BB}=\frac{A_{BB}}{\ell+\ell_*},\hspace{6pt}C_\ell^{EB}=0.
    \end{equation}
    with $\ell_*=10$ and $A_{BB}=0.2$. The relatively large $E$-$B$ asymmetry ensures that any inaccuracy in the estimated mode-coupling matrix will translate into a significant bias in the $B$-mode power spectrum. These simulations were generated using the {\tt HEALPix} pixelisation scheme \citep{astro-ph/0409513} with resolution parameter $N_{\rm side}=1024$, corresponding to pixels of $\sim3.4'$.

    In the first set of simulations we make use of a base binary mask constructed by removing the regions of the sky most contaminated by Milky Way dust and stars (see \cite{1607.01182} for further details). This mask is then apodised with a $1^\circ$ Gaussian kernel. Starting from this mask (labelled $m$ below), we construct the full weighting matrix as
    \begin{align}\nonumber
      v_{11}(\nv)=(1&+\delta_m)\,m(\nv),\hspace{12pt}v_{22}(\nv)=(1-\delta_m)\,m(\nv)\\
      &v_{12}(\nv)=r_m\,\sqrt{v_{11}(\nv)v_{22}(\nv)}.
    \end{align}
    The parameter $\delta_m$ quantifies the level of asymmetry between the inverse noise variance in the two components of the field (if the weight matrix is to be understood as corresponding to an inverse-variance weighting scheme), while $r_m$ controls the level of correlation between them. This parametrisation ensures that the trace of the matrix, which fully determines the isotropic component $\bar{v}$, stays constant as we vary $\delta_m$ and $r_m$.
    
    Fig. \ref{fig:wmat} shows the anisotropic masks for $\delta_m=0.9$ and $r_m=0.5$, as well as the rows of the mode-coupling matrix that control the level of leakage from the $EE$ power spectrum to the $EE$, $BB$, and $EB$ spectra at $\ell=30$ (from left to right) for different values of $\delta_m$ and $r_m$. We see that both $\delta_m$ and $r_m$ have a commensurate impact in terms of $EE\rightarrow BB$ leakage, while the asymmetry between spin components (controlled by $\delta_m$) seems to dominate the parity-odd $EE\rightarrow EB$ leakage term. Fixing $\delta_m=0.9$ and $r_m=0.5$, corresponding to a very strongly anisotropic weighting, we generate 100 simulations as described above. We generate two versions of each simulation, one masked using the base mask $m$ (corresponding to isotropic weighting), and another one employing the anisotropic weight matrix $v_{ij}$. We then estimate the power spectra of the two weighted maps using the standard pseudo-$C_\ell$ and its anisotropic version, respectively. To reduce the statistical noise of the results, we bin the measured power spectra into bandpowers of width $\Delta\ell=16$. Fig. \ref{fig:clval1} shows the input power spectrum (blue lines), together with the mean of the power spectra estimated from the isotropically-weighted simulations (black markers) and the anisotropically-weighted simulations (red markers). Results are shown for the $EE$ power spectrum (solid line and circular markers), and for the $BB$ spectrum (dashed line and crosses). The bottom panel in the figure shows the residuals with respect to the theory prediction as a fraction of the statistical uncertainties. We find that the standard and anisotropic estimators are both able to obtain unbiased measurements of the power spectrum. For illustrative purposes, the yellow dashed line in the figure shows the $BB$ power spectrum found by analysing the anisotropically-weighted simulations using the standard pseudo-$C_\ell$ estimator, which neglects the anisotropic components of the mask. The estimated power spectrum is strongly biased, by a factor $\sim3.5$ with respect to the true input spectrum.
    \begin{figure*}
      \centering
      \includegraphics[width=0.49\textwidth]{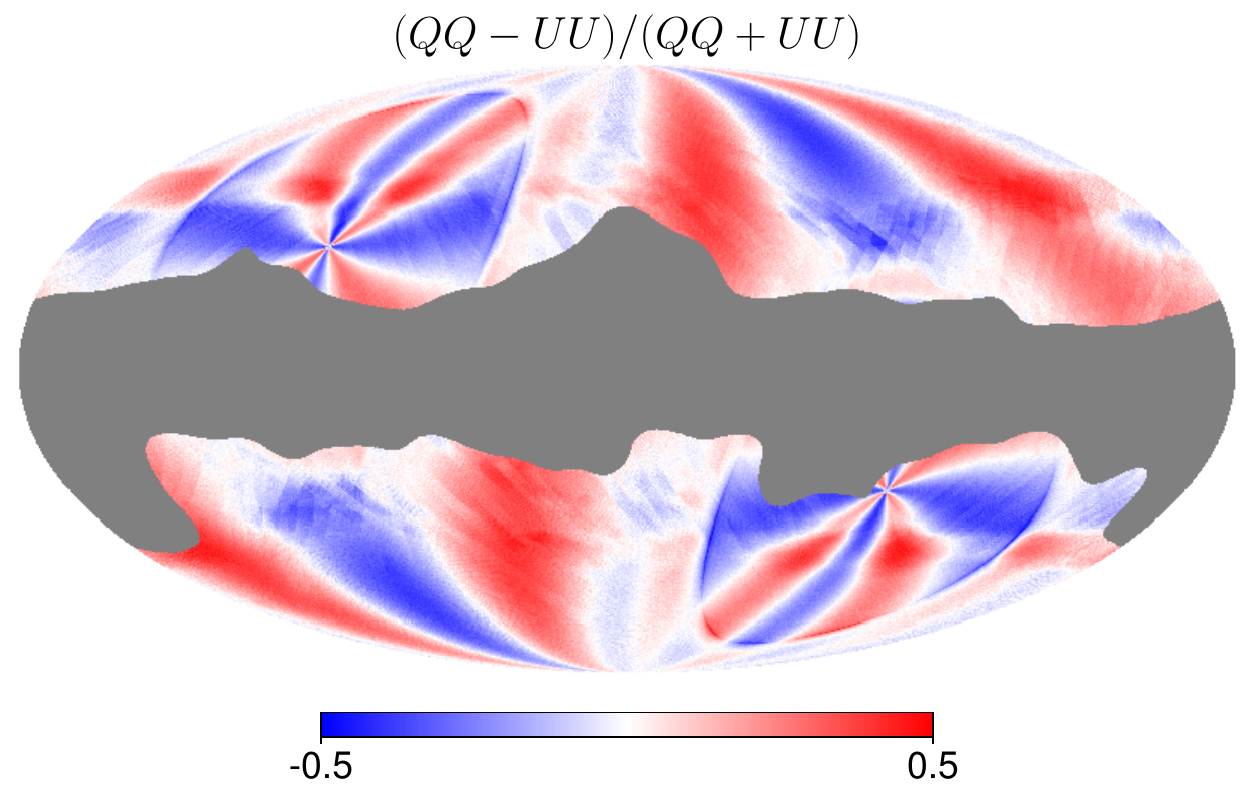}
      \includegraphics[width=0.49\textwidth]{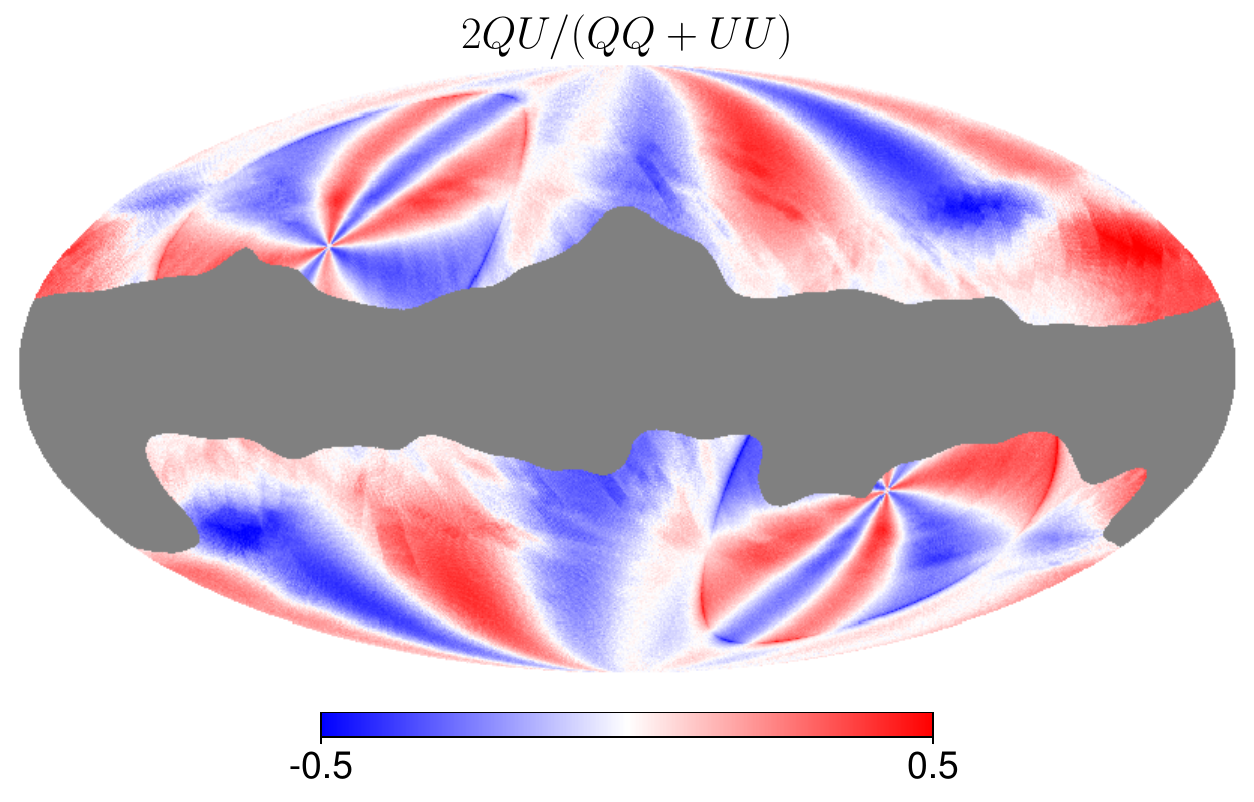}
      \caption{Normalised anisotropic components of the polarisation covariance matrix for the WMAP Ka band map over the footprint allowed by the \planck 60\% Galactic mask.}
      \label{fig:wij_wmap}
    \end{figure*}
    \begin{figure}
      \centering
      \includegraphics[width=0.49\textwidth]{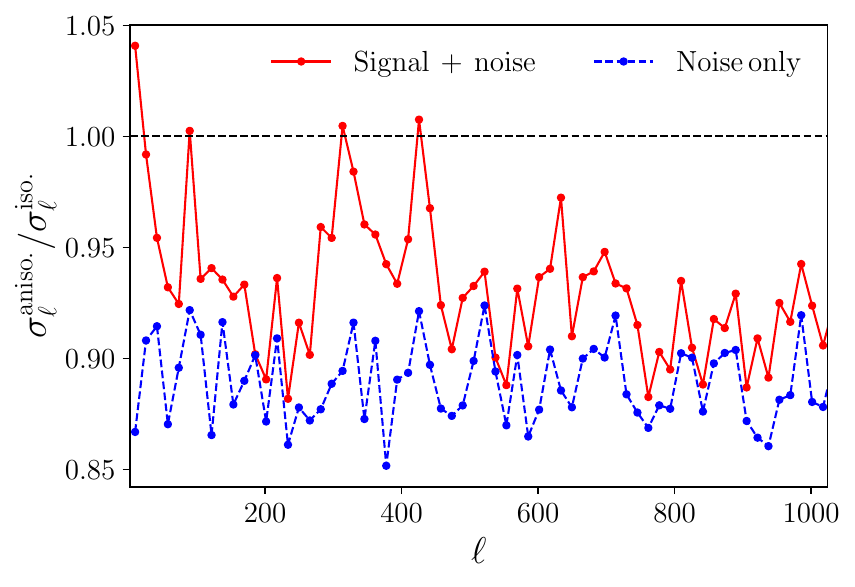}
      \caption{Ratio between the $E$-mode power spectrum uncertainties obtained from maps with anisotropic noise using anisotropic weights that track inverse noise covariance, and the uncertainties obtained using an isotropic mask. The dashed blue and solid red lines show the results for simulations containing only noise and both signal and noise, respectively. The use of anisotropic weights leads to a mild $\sim10\%$ reduction in the statistical uncertainties.}
      \label{fig:cl_error_ratio}
    \end{figure}

    Our second set of validation simulations aims to explore the impact of applying different masks to the two spin components of a field. This mimics a potential use case in which, for example, Galactic contamination may be significantly stronger in particular areas of the sky for one of the two polarisation components. In this case we construct two sky masks using the \planck 60\% and 40\% Galactic masks\footnote{These masks are publicly available at the \planck Legacy Archive \url{https://pla.esac.esa.int/}.} We apply these masks to the $a_1$ and $a_2$ components of the simulated sky maps, respectively (i.e. the components of the weighting matrix in this case are $v_{11}=m_{60\%}$, $v_{22}=m_{40\%}$, and $v_{12}=0$). As in the previous case, we calculate the power spectra of simulations masked using only the isotropic component of the weight matrix, and for anisotropically-weighted maps. The result of this validation exercise is shown in Fig. \ref{fig:clval2}, using the same color scheme as Fig. \ref{fig:clval1}. As before, the anisotropic pseudo-$C_\ell$ method is able to recover an unbiased estimate of the power spectrum. Furthermore, although $v_{11}$ and $v_{22}$ coincide over $\sim80\%$ of the sky, ignoring the mismatch between them leads to a severely biased estimate of the $B$-mode power spectrum, as shown by the dashed yellow line.

  \subsection{Realistic anisotropic weights}\label{ssec:val.wmap}
    In the previous section we tested the anisotropic pseudo-$C_\ell$ estimator against simulations generated using highly-anisotropic weighting schemes, which is appropriate in order to stress-test our implementation. In this section, we instead aim to quantify the potential benefits of using the anisotropic estimator for datasets with realistic levels of anisotropic noise.

    Specifically, we make use of the CMB polarisation noise covariance matrices made available with the 5$^{\rm th}$ data release of the Wilkinson Micowave Anisotropy Probe \citep[WMAP, ][]{1212.5225}\footnote{The data is available on the Legacy Archive for Microwave Background Data Analysis \url{https://lambda.gsfc.nasa.gov/}.} Specifically, we use the $QQ$, $UU$, and $QU$ components of the noise covariance matrix for the Ka band, centered at 33 GHz. Fig. \ref{fig:wij_wmap} shows the anisotropic components, $(QQ-UU)/(QQ+UU)$ and $2QU/(QQ+UU)$, of this noise covariance, which reach levels of up to $\sim50\%$.

    Using this noise covariance, we generate noisy realisations of the CMB sky as follows. We first generate signal-only, full-sky Gaussian realisations following the CMB $EE$ and $BB$ power spectra corresponding to the best-fit $\Lambda$CDM parameters found by \planck. We then generate statistically-isotropic white-noise realisations roughly corresponding to the noise level achieved by the \planck satellite at 44 GHz (as we will see, the exact noise level does not affect the results presented below). Finally, we induce anisotropy in these noise realisations, following the structure of the WMAP covariance, by multiplying them by the Cholesky decomposition of the following matrix:
    \begin{equation}\nonumber
      \left(\begin{array}{cc}
         1+\delta_1  & \delta_2 \\
         \delta_2  & 1-\delta_1
      \end{array}\right),\hspace{12pt}\delta_1\equiv\frac{QQ-UU}{QQ+UU},\hspace{12pt}\delta_2\equiv\frac{2QU}{QQ+UU}.
    \end{equation}
    We then generate the noisy sky realisation by adding these signal and noise contributions.

    The power spectrum of each simulation is estimated in two different ways. In the first case we mask the simulated maps using a simple isotropic mask, given by the $60\%$ \planck Galactic mask smoothed with a $0.5^\circ$ Gaussian kernel. In the second case, we apply an anisotropic weighting matrix to the simulations, constructed by multiplying the sky mask we just described with the pixel-level inverse noise covariance. Thus, although we use the same sky area to estimate the power spectra in both simulations, the anisotropy in the noise is only optimally accounted for in the second set of simulations. We generate 500 such simulations at resolution $N_{\rm side}=512$. The analysis is repeated for simulations containing only noise an both signal and noise.
    
    The main result from this analysis is shown in Fig. \ref{fig:cl_error_ratio}. The figure shows the standard deviation of the power spectra measured from the anisotropically-weighted simulations as a fraction of the same variance calculated from the simulations with isotropic weights. The results are shown for the noise-only simulations (blue dashed line) and for the simulations containing both signal and noise (red solid line). In the noise-only case, we find that accounting for the anisotropy in the noise leads to a $10\%$ reduction in the statistical uncertainties. For simulations including signal, this improvement is reduced on the larger, signal-dominated scales, as could be expected. This level of improvement is perhaps surprisingly mild (considering the relatively large level of anisotropy implied by Fig. \ref{fig:wij_wmap}). We can therefore expect that, in terms of achieving optimal statistical uncertainties, anisotropic weighting will only be relevant for noise-dominated data exhibiting very significant levels of anisotropy. Nevertheless, as discussed in Section \ref{ssec:meth.genpcl}, this error reduction is achieved at virtually no computational cost in the power spectrum estimator.

\section{Conclusion}\label{sec:conc}
  The standard pseudo-$C_\ell$ estimator, used ubiquitously in analyses of projected anisotropies in cosmology and astrophysics, when applied to non-scalar fields, assumes that all spin components are weighted or masked in the same way. This procedure is sub-optimal for fields with anisotropic noise properties (e.g. CMB polarisation observations with large differences in the sensitivities of orthogonal detector pairs, or imaging artifacts in galaxy shape measurements for weak lensing), or when different sky masks must be applied to different spin components. In this paper we have presented a generalisation of this standard approach to account for this type of component-wise or anisotropic weighting.
  
  Our methodology has consisted of separating the weighting matrix into its scalar and spin-$2s$ components, with the latter incorporating all the purely anisotropic noise properties. The expressions derived for the pseudo-$C_\ell$ mode-coupling coefficients involve additional coupling between parity-odd and parity-even fields (mediated by the $B$-mode components of the spin-$2s$ weights), as well as modifications to the standard coupling coefficients involving spectra with the same parity. Although these expressions are mathematically more complex, their computational complexity is comparable to the standard pseudo-$C_\ell$ estimator. Thus, it is possible to incorporate anisotropic weights in the estimator without incurring a significant computational cost. We have validated our implementation of this generalised estimator on simulations with different types of anisotropic weights, including both toy models and realistic levels of anisotropy based on real data. 
  
  Although the gains in terms of sensitivity may be modest for realistic levels of noise anisotropy, they can be achieved at almost no additional cost, and the generalised estimator is vital to obtain unbiased power spectra when different masks must be applied to different spin components. In this paper we have only introduced the generalisation of the bare power spectrum estimator. Further work will be needed to extend the support for anisotropic weights to the calculation of pseudo-$C_\ell$ covariances \citep{astro-ph/0307515,1906.11765}, $B$-mode purification \citep{astro-ph/0511629,0903.2350}, contaminant deprojection \citep{1609.03577,1809.09603}, the application to discrete point processes \citep{2312.12285,2407.21013}, and the applicability of accelerating approximations \citep{2010.14344}. Such extensions may be the subject of future work.

\section*{Acknowledgements}
  I would like to thank Graeme Addison, Will Coulton, Thibaut Louis, and Nicolas Tessore for useful comments and discussions, as well as the South East Madrid Club de Lectura for providing the inspiring environment that initiated this work. I also acknowledge support from the Beecroft Trust.

\bibliography{main}

\appendix
\section{Simplified expressions}\label{app:simpler}
  A more succinct version of the results presented in Eq. \ref{eq:mcm_gen} can be written in terms of the relation between the power spectra $C_\ell^{ab}$ and $C^{a\bar{b}}_\ell$ and their pseudo-$C_\ell$ estimates before separating them into their $E$- and $B$-mode contributions. In the presence anisotropic weights, equations \ref{eq:pcl_ab} and \ref{eq:pcl_abbar} change into
  \begin{align}\nonumber
    \langle\tilde{C}_\ell^{ab}\rangle=M^{\bar{v}\bar{w}}_{\ell\ell'}C^{ab}_{\ell'}-M^{\bar{v}\spt{w}}_{\ell\ell'}C^{a\bar{b}}_{\ell'}-M^{\spt{v}\bar{w}}_{\ell\ell'}C^{\bar{a}b}_{\ell'}+M^{\spt{v}\spt{w}}_{\ell\ell'}C^{\bar{a}\bar{b}}_{\ell'},\\
    \langle\tilde{C}_\ell^{a\bar{b}}\rangle=\bar{M}^{\bar{v}\bar{w}}_{\ell\ell'}C^{a\bar{b}}_{\ell'}-\bar{M}^{\bar{v}\bar{\spt{w}}}_{\ell\ell'}C^{ab}_{\ell'}-\bar{M}^{\spt{v}\bar{w}}_{\ell\ell'}C^{\bar{a}\bar{b}}_{\ell'}+\bar{M}^{\spt{v}\bar{\spt{w}}}_{\ell\ell'}C^{\bar{a}b}_{\ell'},
  \end{align}
  where we have defined the coupling matrices
  \begin{align}
    M^{\bar{v}\spt{w}}_{\ell\ell'}\equiv(-1)^{s_a}\sum_{\ell''}\frac{(2\ell'+1)(2\ell''+1)}{4\pi}\tilde{C}^{\bar{v}\spt{w}}_{\ell''}\wtj{\ell}{\ell'}{\ell''}{s_a}{-s_a}{0}\wtj{\ell}{\ell'}{\ell''}{s_b}{s_b}{-2s_b},\\
    M^{\spt{v}\bar{w}}_{\ell\ell'}\equiv(-1)^{s_b}\sum_{\ell''}\frac{(2\ell'+1)(2\ell''+1)}{4\pi}\tilde{C}^{\spt{v}\bar{w}}_{\ell''}\wtj{\ell}{\ell'}{\ell''}{s_a}{s_a}{-2s_a}\wtj{\ell}{\ell'}{\ell''}{s_b}{-s_b}{0},\\
    M^{\spt{v}\spt{w}}_{\ell\ell'}\equiv\sum_{\ell''}\frac{(2\ell'+1)(2\ell''+1)}{4\pi}\tilde{C}^{\spt{v}\spt{w}}_{\ell''}\wtj{\ell}{\ell'}{\ell''}{s_a}{s_a}{-2s_a}\wtj{\ell}{\ell'}{\ell''}{s_b}{s_b}{-2s_b}.
  \end{align}
  As in Eq. \ref{eq:mcm_std_b}, any barred coupling matrix (e.g. $\bar{M}^{\bar{v}\spt{w}}_{\ell\ell'}$) includes a factor $(-1)^{\sum_\ell}$ inside the sum. Finally, the pseudo-$C_\ell$s of the masks entering these expressions are the complex-valued power spectra, introduced in e.g. Eq. \ref{eq:pcldef_gen} for the fields themselves, applied to the spin-$0$ and spin-$2s$ masks.

\end{document}